\documentclass[AMA,STIX1COL]{WileyNJD-v2}
\newcommand{\beginsupplement}{%
        \setcounter{table}{0}
        \renewcommand{\thetable}{S\arabic{table}}%
        \setcounter{figure}{0}
        \renewcommand{\thefigure}{S\arabic{figure}}%
     }

\articletype{Article Type}%

\received{26 April 2016}
\revised{6 June 2016}
\accepted{6 June 2016}

\begin{document}

\title{Generalized Estimating Equations for Hearing Loss Data with Specified Correlation Structures}

\author[1]{Zhuoran Wei} \author[2]{Hanbing Zhu} \author[3,4]{Sharon Curhan} \author[3,4,5,6]{Gary Curhan}
\author[1,3,4,6]{Molin Wang*}

\authormark{Wei and Zhu \textsc{et al}}

\address[1]{\orgdiv{Department of Biostatistics}, \orgname{Harvard T.H. Chan School of Public Health}, \orgaddress{Boston, \state{MA}, \country{USA}}}

\address[2]{\orgdiv{Key Laboratory of Advanced Theory and Application in Statistics and Data Science-MOE, School of Statistics}, \orgname{East China Normal University}, \orgaddress{Shanghai, \country{China}}}

\address[3]{\orgname{Harvard Medical School}, \orgaddress{Boston, \state{MA}, \country{USA}}}

\address[4]{\orgdiv{Channing Division of Network Medicine, Department of Medicine}, \orgname{Brigham and Women’s Hospital}, \orgaddress{Boston, \state{MA}, \country{USA}}}

\address[5]{\orgdiv{Division of Renal Medicine, Department of Medicine}, \orgname{Brigham and Women’s Hospital}, \orgaddress{Boston, \state{MA}, \country{USA}}}

\address[6]{\orgdiv{Department of Epidemiology}, \orgname{Harvard T.H. Chan School of Public Health}, \orgaddress{Boston, \state{MA}, \country{USA}}}

\corres{*Molin Wang, Department of Epidemiology, Harvard T.H. Chan School of Public Health, Boston, MA, USA.
\email{stmow@channing.harvard.edu}}

\abstract[Summary]{Due to the nature of pure-tone audiometry test, hearing loss data often has a complicated correlation structure. Generalized estimating equation (GEE) is commonly used to investigate the association between exposures and hearing loss, because it is robust to misspecification of the correlation matrix. However, this robustness typically entails a moderate loss of estimation efficiency in finite samples. This paper proposes to model the correlation coefficients and use second-order generalized estimating equations to estimate the correlation parameters. In simulation studies, we assessed the finite sample performance of our proposed method and compared it with other methods, such as GEE with independent, exchangeable and unstructured correlation structures. Our method achieves an efficiency gain which is larger for the coefficients of the covariates corresponding to the within-cluster variation (e.g., ear-level covariates) than the coefficients of cluster-level covariates. The efficiency gain is also more pronounced when the within-cluster correlations are moderate to strong, or when comparing to GEE with an unstructured correlation structure. As a real-world example, we applied the proposed method to data from the Audiology Assessment Arm of the Conservation of Hearing Study, and studied the association between a dietary adherence score and hearing loss. 
}


\maketitle

\section{Introduction}\label{sec1}

Hearing loss is a common and growing disease burden currently affecting one in five people in the world. By 2050, the number of people living with disabling hearing loss is expected to exceed 700 million \cite{GDB2019}. Besides societal prejudice and marginalization, auditory defects could lead to poor psycho-social well-being, low economic income, unsatisfactory personal achievement and so forth \cite{davis2019}. Globally, about 60\% of the hearing loss in children could be prevented \cite{who}, so there are salient benefits to investigate potential causes for defective hearing. Conservation of Hearing Study (CHEARS) is a nationwide cohort study aimed to reveal the relationship between hearing loss and various medical, dietary and other lifestyle factors, and possibly shed light on the risk factors useful in preventing hearing loss onset and progression. One data source of CHEARS is the ongoing Nurses' Health Study II (NHS II), which started in 1989 with 116,430 participants. A subset of 3,136 NHS II participants composed the Audiology Assessment Arm (AAA), and underwent baseline and 3-year follow-up pure-tone audiometry tests. This audiometric assessment was used to detect longitudinal changes in the patient's hearing thresholds for both ears at 7 frequencies \cite{chears}. Data from CHEARS-AAA will be used as a real-world application of the method we propose.

\indent Pure-tone audiometry is the gold-standard test for diagnosing hearing loss and determining its type, degree and configuration. It measures audibility thresholds at frequencies ranging from 250 Hz to 8000 Hz \cite{jilljin}. Hearing loss can be diagnosed based on either ear’s relative change of hearing threshold at different frequencies over time \cite{moore2022}. The outcomes (i.e. change of hearing threshold) from two ears of a participant are correlated, and the outcomes at different frequencies from the same ear are also correlated. This multilayered correlation structure of the audiometric data post challenges to statistical analyses.

\indent Most of the existing hearing loss studies use single-ear methods, such as the worse-ear method and better-ear method, which utilize hearing measurements from only one ear of the participant, for example the worse ear for the worse-ear method and the better ear for the better-ear method. Another commonly used method is the average-ear method, in which the outcome is the mean of hearing threshold changes from both ears. These methods could lead to bias and inefficient estimates. Sheng et al. (2022) describes a linear mixed-effect model which uses the both-ear method to account for ear-level correlation in the audiometric data \cite{sheng}. They illustrate that, with only participant-level covariates in the model, the single-ear methods are unbiased but typically less efficient than the average- and both-ear method. However, when ear-level covariates are included in the model, the single-ear methods could introduce bias, and the average-ear method is unbiased while less efficient than the both-ear method. 

\indent To handle the correlated pure-tone audiometry data, generalized estimating equation (GEE) could be used to estimate the association between exposure and hearing loss \cite{liangezeger}. GEE outperforms generalized mixed effect models (GLMM) in some aspects. GLMM requires assumptions on the conditional distribution of the outcome and the distribution of the random effects. On the other hand, GEE is a semi-parametric method, which does not need assumptions about the distribution of the data \cite{geeornot}.

\indent Most existing hearing loss studies assumed an independent or exchangeable working correlation matrix in GEE \cite{rigter}. Chen et al. (2022) compare the both-ear and single-ear method in the analysis of binary audiometric data, using GEE with exchangeable and unstructured correlation matrix \cite{chenchen}. They reach similar conclusion as Sheng et al. (2022). For the GEE method, as long as the mean model is correctly specified, the point estimates and standard errors are consistent even with misspecified correlation structure and heteroskedastic residuals \cite{chenchen}\cite{fitzmaurice}. However, using misspecified correlation matrix could lead to larger finite sample bias, especially if used along with single-ear method. Moreover, assuming an independence correlation matrix may result in a loss of efficiency, especially when estimating the coefficient for covariate that leads to ``within-cluster variation'' \cite{fitzmaurice}. 

\indent In this paper, we propose a statistical method that models the correlation matrix for GEE and takes account of special characteristics of pure-tone audiometry data. The proposed method could improve the efficiency of regression coefficient estimation in studying the association between exposures and hearing loss, which is important as the sample size of pure-tone audiometry data is usually not large due to the resource- and time-consuming nature of the pure-tone audiometry tests. In Section 2, we introduce the audiometric data and issues ensued in modelling the association between hearing loss and exposures. In Section 3, we presents our proposed estimation methods. In Section 4, we conduct simulation studies to compare our methods with other commonly used methods. In Section 5, we apply our methods to estimate the association between hearing loss and the Dietary Approaches to Stop Hypertension (DASH) diet adherence scores using CHEARS-AAA data as a real-world illustrative example. 

\section{Model and Notation}\label{sec2}

We define the outcome variable, hearing loss, as the change of hearing threshold between measurements at baseline and at the end of study.  Assume that there are $n$ participants in the study. We use $Y_{ijq}$ to denote the outcome for the $j$th ear of the $i$th participant at the $q$th frequency ($i = 1, \ldots, n$; $j = 1,2$; $q = 1, \ldots Q$). Each participant is a cluster, within which hearing threshold data at different frequencies for each ear form the ear level sub-cluster. In addition, hearing threshold data from different ears of the same participant measured at the same frequency form the frequency level sub-cluster. Therefore, the ear and frequency sub-clusters are nested within participant cluster, but not nested within each other. 

\indent We assume the model for the conditional mean  $\mu_{ijq} := E(Y_{ijq}|\boldsymbol{X}_{ijq} )$ is \begin{equation} \label{equ:1} g(\mu_{ijq})=\mathbf{X}_{ijq}^{\top}\boldsymbol{\beta} =\beta_{0}+\sum_{q=2}^{Q}\left(\beta_{q}\mathbf{I}(\text{frequency}=q)\right) + \mathbf{X}_{ijq}^{\top}\boldsymbol{\beta}^{(1)}+\sum_{q=2}^{Q}\mathbf{X}_{ijq}^{\top}\boldsymbol{\beta}^{(2)}_{q}\mathbf{I}(\text{frequency}=q),\end{equation} where $g(\cdot)$ is a known link function, $\mathbf{I}(\text{frequency}=q)$ is the indicator variable for frequency $q$, and $g(\mu_{ijq})$ is associated with a $p$-dimensional covariate vector $\mathbf{X}_{ijq} = [\mathbf{\tilde{X}}_{ijq}, \mathbf{\check{X}}_{ijq}]^\top$, which includes participant level exposures and potential confounders $\mathbf{\tilde{X}}_{ijq}$, (e.g., $\mathbf{\tilde{X}}_{i11} = \mathbf{\tilde{X}}_{i22}$), and ear- and frequency-level exposures and potential confounders $\mathbf{\check{X}}_{ijq}$, (e.g., $\mathbf{\check{X}}_{i11} \neq \mathbf{\check{X}}_{i21}$ or $\mathbf{\check{X}}_{i11} \neq \mathbf{\check{X}}_{i12}$). In CHEARS-AAA example, the participant-level exposure of interest is DASH, and potential participant-level confounders include body mass index (BMI), aspirin intake, and so on. One potential ear- and frequency-level covariate is the hearing threshold at baseline for each ear at a given frequency. In Model \eqref{equ:1} above, we focus on the identity link $g(\mu_{ijq}) = \mu_{ijq}$ in the simulation study and real data example. However, the methods also apply to non-identity link. The scalar parameter $\beta_q, q = 2, \ldots, Q$, is the covariate corresponding to the $q$th frequency. The $p$-dimensional vectors $\boldsymbol{\beta}^{(1)}$ and $\boldsymbol{\beta}^{(2)}_q$ are parameters for covariates and their interactions with frequency $q$ respectively. Thus, $\beta_0$, $\beta_q$, $\boldsymbol{\beta}^{(1)}$ and $\boldsymbol{\beta}^{(2)}_q, q = 2, \ldots, Q,$ are all elements of the coefficient vector $\boldsymbol{\beta}$.  Without further specification, all vectors are column vectors throughout this paper. Superscript $T$ represents transpose. We assume Model \ref{equ:1} with only linear terms for notational simplicity. The methods described below work for more flexible models, for example, models with more interaction terms, and nonlinear terms. 

\indent We use GEE to estimate the regression coefficients in Model \ref{equ:1}, as GEE takes into account the correlation in outcome measurements within the same cluster. Let $\mathbf{Y}_i = (Y_{i11}, Y_{i21}, Y_{i12}, \ldots$, $Y_{i1Q}, Y_{i2Q})^{\top}$ and $\boldsymbol{\mu}_i = \{ E(\mathbf{Y}_{i}|\mathbf{X}_{ijq}); j = 1,2; q = 1,\ldots, Q\}$. In audiometric threshold data, we often view each participant as one cluster. We can write the GEE as 
\begin{equation} \label{equ:2}
  \mathbf{U}(\boldsymbol{\beta}) = \sum^{n}_{i = 1}\mathbf{D}_i^{\top}\mathbf{V}_i^{-1} (\mathbf{Y}_i - \boldsymbol{\mu}_i)  =\mathbf{0},
\end{equation}
where $\mathbf{D}_{i}$ is the first derivative of $\boldsymbol{\mu}_i$ with respect to the coefficients $\boldsymbol{\beta}$, and $\mathbf{V}_i$ is the $N \times N$ working covariance matrix of $\mathbf{Y}_i$, $N = 2Q$. If observations within participants are independent, the working covariance of measurements from the $i$th participant can be the true variance, expressed as
\begin{equation}\label{equ:vaia}\mathbf{V}_i = \text{Cov}(\mathbf{Y}_i) = \mathbf{A}_i =\mathbf{A}_i^{1/2}\mathbf{I}\mathbf{A}_i^{1/2},\end{equation}
where $\mathbf{I}$ is the $N \times N$ identity matrix, $\mathbf{A}_i = \text{diag}\big{(}\upsilon(\mu_{i11}), \upsilon(\mu_{i21}), \ldots, \upsilon(\mu_{i1Q}), \upsilon(\mu_{i2Q})\big{)}$, which is a diagonal matrix containing variances of the elements of $\mathbf{Y}_i$, and $v(\cdot)$ is the variance function, Var($Y_{ijq}$) = $v(Y_{ijq})$. To handle correlated outcomes within each participant, we can replace the identity matrix with a more general correlation
matrix, so the working covariance matrix can be re-written as 
\begin{equation} \label{equ:vara} \mathbf{V}_i = \mathbf{A}_i^{1/2} \mathbf{R}_i(\boldsymbol{\alpha}) \mathbf{A}_i^{1/2},\end{equation} where $\mathbf{R}_i(\boldsymbol{\alpha})$ is  a ``working'' correlation matrix of $\mathbf{Y}_i$, possibly dependent on a parameter vector $\boldsymbol{\alpha}$. Thus, we can re-write the estimating equation \eqref{equ:2} as
\begin{equation} \label{equ:3}
\mathbf{U}(\boldsymbol{\beta}, \boldsymbol{\alpha}) = \sum_{i=1}^{n} \mathbf{D}_i^{\top}\big\{ \mathbf{A}_i^{1/2}\mathbf{R}_i(\boldsymbol{\alpha}) \mathbf{A}_i^{1/2}\big\}^{-1}(\mathbf{Y}_i- \boldsymbol{\mu}_i) = \mathbf{0}.
\end{equation}

Even under misspecification of the working covariance matrix, $V$, the GEE leads to consistent $\boldsymbol{\beta}$-estimators. However, if $V$ is closer to the true covariance matrix of $Y_i$, the $\boldsymbol{\beta}$-estimators will typically be more efficient \cite{fitzmaurice}. Because the hearing loss data has a special correlation structure, it may be less efficient to use the unstructured or pre-existing types of working correlation matrix (e.g. exchangeable, auto-regressive, etc.) as in a standard GEE analysis. Hence, we specify a working correlation matrix for the audiometric data collected in a typical hearing loss study; one example is the CHEARS-AAA data described in the introduction section. Let $\rho_{p_1p_2}$ be the $(p_1, p_2)$th element of $\mathbf{R}_i(\boldsymbol{\alpha})$, $\{p_1, p_2 = 1, \ldots, N\}$. Hence, $\rho_{p_1p_2}$ is the correlation coefficient for the two outcome measurements $\mathbf{Y}_{ip_1}$ and $\mathbf{Y}_{ip_2}$ from the $i$th participant, where $Y_{ip}$ denotes the $p$th element of $Y_i$. 

\indent For audiometric data, we propose the following correlation model:
\begin{equation*}
\rho_{p_{1} p_{2}}=1-\alpha_{0} \alpha_{e}^{\mathbf{I}\big(j\left(p_{I}\right)=j\left(p_{2}\right)\big)} \alpha_{f}^{\mathbf{I}\big(q\left(p_{I}\right)=q\left(p_{2}\right)\big)},
\end{equation*}
where $j(p_1)$ and $j(p_2)$ index the ears of the outcomes $Y_{ip_1}$ and $Y_{ip_2}$, respectively, and $q(p_1)$ and $q(p_2)$ represent the frequencies for the outcomes $Y_{ip_1}$ and $Y_{ip_2}$, respectively. Accordingly, the within-person correlation coefficient is $1-\alpha_{0}$ for measurements from different ears, different frequencies; $1-\alpha_{0} \alpha_{f}$ for measurements from different ears at the same frequency; and $1-\alpha_{0} \alpha_{e}$ for the measurements from the same ear at different frequencies. Table 1 shows the correlation coefficients in \eqref{equ:3} between two measurements of the same participants, using the $0.5 \mathrm{kHz}$ and $1 \mathrm{kHz}$ frequencies as examples. The restrictions such as $0 \leq \alpha_{0} \leq 1,0 \leq \alpha_{e} \leq 1,$ $0 \leq \alpha_{f}  \leq 1$ will need to be taken care in estimating the parameters $\boldsymbol{\alpha}=\left(\alpha_{0}, \alpha_{f}, \alpha_{e}\right)^{\top},$ for example, by using appropriate transformation such as $\textup{exp}(\cdot)/\{1+\textup{exp}(\cdot)\}$. 

\begin{table}[htbp]
\caption{Within-person correlation coefficients}
\centering
\begin{tabular}{c|cccc}
\toprule
& left,0.5kHz  & right,0.5kHz & left,1kHz & right,1kHz \\
\hline
left,0.5kHz & 1 & $1-\alpha_{0}\alpha_{f}$ & $1-\alpha_{0} \alpha_{e}$ & $1-\alpha_{0}$ \\
right,0.5kHz & $1-\alpha_{0} \alpha_{f}$ & 1 & $1-\alpha_{0}$  & $1-\alpha_{0} \alpha_{e}$ \\
left,1kHz & $1-\alpha_{0} \alpha_{e}$ &  $1-\alpha_{0}$ & 1 & $1-\alpha_{0} \alpha_{f}$ \\
right,1kHz & $1-\alpha_{0}$ & $1-\alpha_{0} \alpha_{e}$ & $1-\alpha_{0} \alpha_{f}$ & 1\\ 
\bottomrule
\end{tabular}
    \text{For example, ``left,0.5kHz'' represents left ear measurement for frequency $0.5 \mathrm{kHz}$.}
\end{table}

\section{Estimation Methods}\label{sec3}

\noindent Based on the standard GEE method, the correlation coefficients $\boldsymbol{\rho} = \{ \rho_{p_1p_2}; p_1, p_2 = 1, \ldots, N\}$ could be estimated by methods of moments \cite{liangezeger}. Using an iterative approach, the correlation parameters $\boldsymbol{\rho}^{(r+1)}$ can be estimated given the coefficients $\boldsymbol{\beta}^{(r)}$ at the $r$th iteration: 

$$
\widehat{\rho}^{(r+1)}_{p_1p_2} = \frac{1}{N^{*} - k}\sum_{i = 1}^{n}\sum_{p_1 < p_2}\widehat{R}_{ip_1}^{(r)}\widehat{R}_{ip_2}^{(r)},
$$ where $\widehat{\boldsymbol{\mu}}^{(r)}_i = \mu_i(\widehat{\boldsymbol{\beta}}^{(r)})$, $N^{*} = \frac{nN(N-1)}{2}$, $k$ is the number of covariates in $\boldsymbol{X}$, and the Pearson residuals $\widehat{R}_{ip}^{(r)} = \frac{Y_{ip}-\hat{\mu}^{(r)}_{ip}}{\sqrt{v(\hat{\mu}^{(r)}_{ip})}}$, $p = 1,\ldots, N$. This method produces a consistent estimator for $\boldsymbol{\beta}$, which means $\sqrt{n}(\widehat{\boldsymbol{\beta}} - \boldsymbol{\beta})$ follows an asymptotic multivariate normal distribution with mean 0 and covariance matrix $\Sigma = \lim_{N \to \infty } n\Sigma_0^{-1}\Sigma_{1}\Sigma_{0}^{-1}$, where  
\begin{equation}
\begin{split}
\Sigma_0 &= \sum^{n}_{i = 1}{\mathbf{D}_i^{\top} \mathbf{V}_i^{-1} \mathbf{D}_i }, \\
\Sigma_1 & = \sum^{n}_{i = 1}{ \mathbf{D}_i^{\top} \mathbf{V}_i^{-1} (\mathbf{Y}_i - \boldsymbol{\mu}_i)(\mathbf{Y}_i - \boldsymbol{\mu}_i)^{\top} \mathbf{V}_i^{-1} \mathbf{D}_i}.
\end{split}
\end{equation}

This yields the ``sandwich estimator'' $\hat\Sigma$ of $\Sigma$ when plugging in the consistent estimates for $\boldsymbol{\beta}$, $\boldsymbol{\rho}$ and $\mathbf{V}_i$ \cite{yan2006}. However, this method has several drawbacks. For example, Kalema et al. (2015) states that first order GEE (GEE1) allows a misspecification of the correlation structure, leading to an arbitrary distribution of $\boldsymbol{Y}_i$, which hampers inference on the correlation structure. The iterative approach could be inefficient compared to model-based estimation since an imprecise estimate of $\boldsymbol{\alpha}$ at a given iteration could cause the model fail to converge \cite{kalema}. Most importantly, it cannot implement a pre-specified correlation structure such as the one in Table 1. Alternatively, Prentice and Zhao (1991) \cite{prenticezhao} proposed using second order GEE (GEE2) to estimate the correlation parameters $\boldsymbol{\alpha}$ for the correlation coefficients $\boldsymbol{\rho}$, and thus $\boldsymbol{\theta} = (\boldsymbol{\beta}, \boldsymbol{\alpha})^\top$ are estimated by solving a joint estimating equation $$
\Psi(\boldsymbol{\theta}) = \sum_{i = 1}^{n}\widetilde{\boldsymbol{D}}^{\top}_i\widetilde{\boldsymbol{V}}_i^{-1}\boldsymbol{E}_i = \boldsymbol{0},
$$ where 
$$
\widetilde{\boldsymbol{D}}_i = \frac{\partial( \boldsymbol{\mu}_i, \boldsymbol{\rho}_i)}{\partial( \boldsymbol{\beta},  \boldsymbol{\alpha})}  =\begin{pmatrix} \frac{\partial \boldsymbol{\mu}_i}{\partial \boldsymbol{\beta} } & \boldsymbol{0} \\  \frac{\partial \boldsymbol{\rho}_i}{\partial \boldsymbol{\beta}} & \frac{\partial \boldsymbol{\rho}_i}{\partial \boldsymbol{\alpha} }\end{pmatrix} = 
\begin{pmatrix}
    \widetilde{\boldsymbol{D}}_{i11} & \widetilde{\boldsymbol{D}}_{i12} \\ 
    \widetilde{\boldsymbol{D}}_{i21} & \widetilde{\boldsymbol{D}}_{i22}
\end{pmatrix}, \;
\boldsymbol{E}_i = \begin{pmatrix} 
\boldsymbol{Y}_{i} - \boldsymbol{\mu}_i    \\
\boldsymbol{Z}_i - \boldsymbol{\rho}_i 
\end{pmatrix},\;  Z_{ip_1p_2} = \frac{(Y_{ip_1} - \mu_{ip_1})(Y_{ip_2} - \mu_{ip_2})}{\sqrt{v(\mu_{ip_1})v(\mu_{ip_2})}}, $$
and $\widetilde{\boldsymbol{V}}_i$ is the working covariance matrix for $\boldsymbol{Y}_{i}$ and $\boldsymbol{Z}_i$; for example, $$\widetilde{\boldsymbol{V}}_i = \begin{pmatrix}
    \text{Var}(\boldsymbol{Y}_{i}) & \text{Cov}(\boldsymbol{Y}_{i}, \boldsymbol{Z}_i) \\
   \text{Cov}(\boldsymbol{Z}_i, \boldsymbol{Y}_{i}) & \text{Var}(\boldsymbol{Z}_{i})
\end{pmatrix}= \begin{pmatrix}
    \widetilde{\boldsymbol{V}}_{i11} & \widetilde{\boldsymbol{V}}_{i12} \\
    \widetilde{\boldsymbol{V}}_{i21} & \widetilde{\boldsymbol{V}}_{i22}
\end{pmatrix}.$$ In this formulation, $\boldsymbol{\mu}_i$ are the model mean, parameterized by $\boldsymbol{\beta}_i$, and $\boldsymbol{\rho}_i$ are the correlation coefficients, expressed through $\boldsymbol{\alpha}_i$ \cite{Ziegler}. 

\indent Due to the complex structure and large size of $\widetilde{\boldsymbol{V}}_i$, directly implementing $\widetilde{\boldsymbol{V}}_i$ can be computationally difficult and inefficient. Thus, we adopt the GEE1.5 method instead, which is a combination of GEE1 and GEE2 \cite{prentice1988}, assuming independence between the two sets of estimating equations (i.e. $\text{Cov}(\boldsymbol{Z}_i, \boldsymbol{Y}_{i}) = \boldsymbol{0}$) \cite{kalema}. Specifically, we first set $\mathbf{R}(\boldsymbol{\alpha})$ to be the identity matrix as in expression \eqref{equ:vaia}, and solve equation \eqref{equ:3} $\mathbf{U(\boldsymbol{\beta})} = 0$  for our ``initial guess'' of $\boldsymbol{\beta}$, denoted as $\widehat{\boldsymbol{\beta}}_0$. In the second step, we specify the correlation structure $\mathbf{R}(\boldsymbol{\alpha})$ in expression \eqref{equ:vara} based on Table 1; that is, $\boldsymbol{\alpha}=(\alpha_0,\alpha_e, \alpha_f)$. We obtain estimates for the correlation coefficients $\widehat{\boldsymbol{\alpha}}$ for given $\widehat{\boldsymbol{\beta}}_0$, by solving the second-order estimating equation \cite{prentice1988}:
$$
\sum^{n}_{i=1}\widetilde{\boldsymbol{D}}^{\top}_{i22}\widetilde{\boldsymbol{V}}_{i22}^{-1}(\boldsymbol{Z}_i - \boldsymbol{\rho}_i) = \boldsymbol{0},
$$
where $\widetilde{\boldsymbol{D}}_{i22} = \partial\boldsymbol{\rho}_i / \partial\boldsymbol{\alpha}$, and $\widetilde{\boldsymbol{V}}_{i22}$ are unstructured working covariance matrices for $\boldsymbol{Z}_i$. 
In the third step, we update $\mathbf{R}(\boldsymbol{\alpha})$ by plugging in $\widehat{\boldsymbol{\alpha}}$ obtained in step 2, and solve equation \eqref{equ:3} again, which gives us $\widehat{\boldsymbol{\beta}}$, the final estimate of $\boldsymbol{\beta}$. We can also repeat step 2 and 3 for more iterations but in our simulation studies the final estimate does not change much after the first iteration.  

\section{Simulation studies}

In this section, we investigate the finite sample performance of our proposed method via simulation studies. We generated data from the regression model with both participant level and ear level covariates: \begin{align*} Y_{ijq} & = \beta_0 + \beta_1 \mathbf{I}(q = 2) + \beta_2 X_{i} + \beta_3 X_{i} \mathbf{I}(q = 2) + \theta Z_{ijq} + \varepsilon_{ijq},
\end{align*} where $i = 1, \ldots, n$; $j = 1, 2$; $q = 1, 2$. For each subject $i$, there are four outcome measurements $Y_{ijq}$, which, for example, correspond to the change in hearing threshold from baseline to 3-year follow-up, from both ears, at 2 frequencies in the CHEARS-AAA motivaitng study. The participant-level covariate $X_i$, for example, corresponds to the DASH score in CHEARS-AAA study. We include its interaction with the frequency $q$ used in pure-tone audiometry test. The coefficient parameters are set close to the true values from an analysis of CHEARS-AAA data set: $(\beta_0, \beta_1, \beta_2, \beta_3)^{\top} = (2.0, -0.7, -1.2, 0.9),$ and $\boldsymbol{\varepsilon}_i = (\varepsilon_{i11}, \varepsilon_{i21}, \varepsilon_{i12}, \varepsilon_{i22})^{\top} \sim N(0, \boldsymbol{\Sigma_\varepsilon})$, where 
$$
\boldsymbol{\Sigma_{\varepsilon}}= 
\begin{pmatrix} 
1  & 0.5 & 0.9 & 0.6 \\  0.5  &1   & 0.6  &0.9 \\
0.9 & 0.6 & 2.25 &1.35 \\ 0.6 & 0.9 & 1.35 & 2.25
\end{pmatrix}. $$

\indent The ear-level covariate $Z_{ijq}$ corresponds to each ear's hearing threshold for each frequency level at baseline in CHEARS-AAA. The ear-level coefficient $\boldsymbol{\theta} = ( \theta_{i11}, \theta_{i21}, \theta_{i12}, \theta_{i22})^{\top} \sim N(-0.8, \boldsymbol{\Sigma_{\theta}}) $, where
$$
\boldsymbol{\Sigma_{\theta}}= 
\boldsymbol{A}\begin{pmatrix} 
1  &1-\alpha_0\alpha_{f_1}  & 1-\alpha_0\alpha_{e} & 1-\alpha_0 \\
1-\alpha_0\alpha_{f_1}  &1   & 1-\alpha_0  &1-\alpha_0\alpha_{e} \\
1-\alpha_0\alpha_{e} & 1-\alpha_0 & 1 &   1-\alpha_0\alpha_{f_2} \\
1-\alpha_0 & 1-\alpha_0\alpha_{e} & 1-\alpha_0\alpha_{f_2} & 1
\end{pmatrix}\boldsymbol{A}, \text{ and }
\boldsymbol{A} = \text{diag}(0.8, 1.9, 0.6, 2.5).
$$

\indent For the working correlation matrix, we first adopt very strong pair-wise correlations for $\boldsymbol{R(\alpha)}$ in Scenario 1. Next, we gradually decrease the strength of correlation in Scenario 2, Scenario 3 and Scenario 4, in order to compare the model performance for varying strength of correlation. The moderate to strong pairwise correlations in Scenario 2 and 3 are close to the true situation in CHEARS-AAA data. Table \ref{tbl:scenario} shows the values used in the four scenarios. 

\begin{table}[htbp] 
\caption{Correlation parameters in simulation scenario 1, 2, 3 and 4}
\centering
\begin{tabular}{c|cccc|c}
\toprule
  &  $\alpha_0$  & $\alpha_{f_1}$  & $\alpha_{f_2}$  & $\alpha_e$ & $\boldsymbol{\textbf{R}(\alpha)}$ \\
 \hline 
\makecell{\textbf{Scenario 1} \\ (very strong correlation)} & 0.2 & 0.3 & 0.7 & 0.55 & 
  $\begin{pmatrix} 
1  & 0.94 & 0.89 & 0.80 \\  0.94  &1   & 0.80  &0.89 \\
0.89 & 0.80 & 1 & 0.86 \\ 0.80 & 0.89 & 0.86 & 1
\end{pmatrix}$ \\  \hline 
\makecell{\textbf{Scenario 2} \\ (strong correlation)} & 0.4 & 0.8 & 0.9 & 0.6 & $\begin{pmatrix} 
1  &  0.68 &0.76& 0.60 \\  0.68 & 1 & 0.60 & 0.76 \\
0.76 & 0.60 & 1 &  0.64  \\   0.60 & 0.76 & 0.64 &1
\end{pmatrix}$ \\  \hline 
\makecell{\textbf{Scenario 3} \\ (moderate correlation)} & 0.6 & 0.8&0.9& 0.65 &  $ \begin{pmatrix} 
1  &0.52 &0.61& 0.40  \\ 0.52& 1& 0.40 & 0.61\\
 0.61 & 0.40 &1&0.46  \\ 0.40 & 0.61 & 0.46 & 1
\end{pmatrix}$  \\ \hline 

\makecell{\textbf{Scenario 4} \\ (weak correlation)} & 0.8 & 0.8 & 0.9 & 0.8  & 
  $\begin{pmatrix} 
1  & 0.36 & 0.36 & 0.20  \\ 0.36 & 1 & 0.20 & 0.36  \\
0.36 & 0.20 & 1 &  0.28 \\  0.20  & 0.36 & 0.28 & 1
\end{pmatrix}$ \\  
 
\bottomrule
\end{tabular}
\label{tbl:scenario}
\end{table}

\indent We compare the following five methods along with our proposed method, and all of them utilize GEE. The worse-ear method only uses measurements from the worse-ear as the outcome, $Y_{iq}^w = \text{max}(Y_{i1q}, Y_{i2q}).$ The average-ear method uses the mean of measurements from two ears as the outcome, $Y^a_{iq} = \text{mean}(Y_{i1q}, Y_{i2q})$. The first two methods are the worse-ear and average-ear method respectively, assuming an independent (ind.) correlation matrix. The third, fourth and fifth method implement the both-ear methods, using an independent (ind.), exchangeable (exch.) and unstructured (uns.) correlation matrix respectively. Our proposed method uses both-ear method and models the working correlation matrix as specified in Section 2 above. 

\indent We use four operating characteristics to assess and compare the performance of the six methods: (1) relative bias ($\%$) = $\frac{1}{S}\sum_{s=1}^{S}(\widehat{\beta}^{(s)}-\beta)/\beta\times 100 \%$, where $\widehat{\beta}^{(s)}$ represents the estimated coefficient in the $s$th simulation, $s = 1, \ldots, S$; (2) empirical standard error (ESE), which is the standard deviation of the $S$ coefficient estimates from $S$ simulations; (3) coverage rate (CR) of the 95\% confidence interval (CI), calculated as $\frac{1}{S}\sum^{S}_{s=1} I (LB^{(s)} \leq \beta \leq UB^{(s)})$, where $LB^{(s)}$ and $UB^{(s)}$ denote the estimated upper bound and lower bound of the 95\% CI for the true coefficient $\beta$ in the $s$th simulation replicate; (4) relative efficiency of our proposed method to the other five methods, defined as the ratio of empirical SE's. 

\begin{table}[htbp]
\caption{Simulation Results from Scenario 1 (very strong correlation)}
\centering
\begin{tabular}{c|l|ccc|ccc}
\toprule
\multirow{2}{2.5em}{} & \multirow{2}{5.5em}{\textbf{\makecell{Method}}}  & \multicolumn{3}{c|}{\textbf{Number of participants = 200}} & \multicolumn{3}{c}{\textbf{Number of participants = 500}} \\ 
\cline{3-8} &
& \makecell{Relative \\ Bias (\%)} & ESE &  CR (\%) & \makecell{Relative \\ Bias (\%)} & ESE & CR (\%)\\
\hline 
 & Worse-ear (ind.) & -113.233 & 3.199 & 92.5 & -98.082 & 2.002 & 90.6\\ 
 & Average-ear (ind.) & 2.858 & 3.177 & 95.8 & 0.353 & 1.936 & 94.7 \\
 $\beta_2$ & Both-ear (ind.) & 2.813 & 3.179 & 95.5 & 0.542 & 1.939 & 94.7 \\
& Both-ear (exch.) & 1.932 & 3.191 & 95.3 & 0.629 & 1.947 & 94.7   \\
& Both-ear (uns.) & 14.283 & 4.147 & 95.1 & 1.327 & 2.557 & 94.4 \\
  & Proposed &  1.765 & 3.186 & 95.3  & 0.322 & 1.944 & 94.6\\ 
\hline
 & Worse-ear (ind.) & -136.418 & 4.525 & 95.1 & -87.907 & 2.899 & 94.6 \\  
   & Average-ear (ind.) &  3.427 & 2.189 & 95.2 & 1.893 & 1.426 & 93.7\\ 
$\beta_3$ & Both-ear (ind.) & 3.502 & 2.191 & 95.2 & 1.810 & 1.428 & 93.4\\
 & Both-ear (exch.) & 3.673 & 2.195 & 94.9 & 1.676 & 1.430 & 93.2 \\
  & Both-ear (uns.) &  5.243 & 7.672 & 95.6 & 0.432 & 4.944 & 94.0 \\
& Proposed & 3.584 & 2.190 & 95.1 & 1.764 & 1.426 & 93.1 \\ 

\hline

 & Worse-ear (ind.) & -103.626 & 0.319 & 5.9 & -105.000 & 0.212 & 0.0\\  
 & Average-ear (ind.) & 1.200 & 0.313 & 91.7 & -0.161 & 0.202 & 93.4\\ 
$\theta$   & Both-ear (ind.) & 0.952 & 0.303 & 92.3 & 0.013 & 0.197 & 93.5 \\
 & Both-ear (exch.) & 0.157 & 0.278 & 92.1 & -0.004 & 0.188 & 93.3 \\ 
& Both-ear (uns.) & 9.843 & 1.303 & 94.2 & 3.568 & 0.589 & 95.3  \\
& Proposed &  0.502 & 0.243 & 92.6 & -0.254 & 0.163 & 94.4\\ 

\bottomrule
\end{tabular}
\label{tbl:result}
\begin{tablenotes}
\begin{itemize}
     $\;\;\; ^a$ ``SES'' is the empirical SE.  \\ 
     $\;\;\; ^b$ ``ind.'', ``exch.'', and ``uns'' stand for independence, exchangeable and unstructured correlation matrix. \\
     $\;\; ^c$ The coefficient $\beta_2$, $\beta_3$ and $\theta$ correspond to the participant-level exposure, the interaction between participant-level exposure and frequency, and the ear-level covariate, respectively.  
\end{itemize}
\end{tablenotes}
\end{table}

\begin{table}[htbp]
\caption{Relative efficiency for $\theta$ under varying strength of correlation (Number of participants = 200)}
\centering
\begin{tabular}{l|cccc}
\toprule
  \multirow{2}{5.5em}{\textbf{\makecell{Method}}}  & \multicolumn{4}{c}{\textbf{Relative Efficiency}}  \\ 
\cline{2-5} &
 \makecell{Scenario 1 \\ (very strong)} & \makecell{Scenario 2 \\ (strong)}& \makecell{Scenario 3 \\ (moderate)} & \makecell{Scenario 4 \\ (weak)}\\
\hline

  Average-ear (ind.) & 0.776 & 0.861 & 0.919 & 0.980 \\ 
 
Both-ear (ind.) & 0.802  & 0.888 & 0.937 & 0.980 \\ 

  Both-ear (exch.) & 0.874 & 0.930 & 0.965& 0.984 \\ 

 Both-ear (uns.) &  0.186 & 0.007 & 0.094 & 0.545 \\

\bottomrule
\end{tabular}
\begin{tablenotes}
\begin{itemize}
     $\;\;\; ^a$ Relative efficiency is the ratio of empirical SE of our proposed method to that of other methods. \\
     $\;\;\; ^b$ ``ind.'', ``exch.'', and ``uns'' stand for independence, exchangeable and unstructured correlation matrix. \\
     $\;\;\; ^c$ The coefficient $\theta$ corresponds to the ear-level covariate.  
\end{itemize}
\end{tablenotes}
\end{table}

\indent We generated 1000 replicates ($S=1000$) of 200 and 500 participants. The results for Scenario 1 (very strong correlations) are shown in Table \ref{tbl:result}. For the point estimate of the coefficients corresponding to the cluster (participant) level covariate $\beta_2$ and its interaction with frequency $\beta_3$, worse-ear (ind.) performs poorly as the relative biases are huge compared to other five models. Both-ear (uns.) also produces large relative bias, though smaller that from the worse-ear method. The other four methods all have relative biases nearly smaller than 5\%. According to studies by Sheng et al. (2022) and Chen et al. (2022), if there is an ear-level covariate in the model, the worse-ear method is biased and invalid; and the average-ear method is valid \cite{sheng}\cite{chenchen}. Our results are consistent with previous findings. The other four methods have similar relative biases and coverage rates for $\beta_2$ and $\beta_3$, suggesting that the finite-sample bias of the average-ear method and both-ear methods are comparable when estimating participant level coefficients. Similar results are observed for the coefficient, $\theta$, of the covariate that accounts for within-cluster variation (i.e. ear-level coefficient).

\indent Regarding efficiency in Scenario 1 (very strong correlation), both-ear (uns.) has much larger standard errors than the other five methods since it is estimating more correlation parameters, which leads to profound loss of efficiency. When modelling the working correlations as in Table 1, there is not much efficiency gain for the participant-level covariate coefficient $\beta_2$ and the interaction coefficient $\beta_3$ compared to the GEE estimators based on the independent or exchangeable correlation structures. On the other hand, for the ear-level coefficient $\theta$, the empirical standard errors from the average-ear (ind.) method is 3.3\% larger than those from the both-ear (ind.) method in the sample of 200 participants. This is consistent with the findings of Sheng et al. (2022) \cite{sheng} that when the model includes ear-level covariate, the both-ear method may be more efficient than the average-ear method. In addition, the empirical standard errors of our proposed method are smaller than those from the other five methods. The results for Scenario 2, 3 and 4 are shown in Supplementary Table 1, 2, and 3.

\indent Table 4 presents the relative efficiency for the ear-level covariate $\theta$ of the proposed method compared to other methods under different correlation levels in Scenario 1 to 4, among 200 participants. The relative efficiencies of our proposed method compared to the other valid methods (except the worse-ear method) are all less than 1, which indicates that our proposed method is more efficient in estimating the within-cluster coefficient $\theta$ than the other methods, under varying strengths of correlation. The efficiency gain is more pronounced when comparing to the both-ear method using an unstructured correlation matrix or under a smaller sample size. In addition, compared to the both-ear method based on the exchangeable correlation matrix, our method has the largest efficiency gain (12.6\%) when the within cluster correlation is very strong (Scenario 1); as the strength of correlation decreases from strong to moderate, efficiency gain of our method decreases from 7.0\% to 3.5\%. In Scenario 4 where the correlation is weak, all the methods produce similar standard errors, as the relative efficiencies all exceed 0.9. Our results agree with Fitzmaurice (1995) \cite{fitzmaurice} that for within-cluster covariates, relative efficiency of GEE with correctly specified correlation structure may increase as the within-cluster correlation increases.

\section{Application: CHEARS study}

\indent To illustrate our approach, we use hearing loss data from CHEARS-AAA study and investigate the association between the DASH score and 3-year change in hearing threshold. The regression model we use is 
$$Y_{ijq} = \beta_0 + \beta_1 \mathbf{I}(q = 2) + \beta_2 \mathbf{I}(q = 3) + \boldsymbol{\beta_3}\mathbf{X}_{ijq} + \boldsymbol{\beta_4}\mathbf{X}_{ijq}\mathbf{I}(q = 2) +  \boldsymbol{\beta_5}\mathbf{X}_{ijq}\mathbf{I}(q = 3) + \varepsilon_{ijq},$$


\begin{table}
\caption{Results from CHEARS-AAA study}
\centering
\begin{sideways}
\begin{tabular}{l|cc|cc|cc|cc}

\toprule
 \multirow{2}{3.5em}{\textbf{Covariate} } &\multicolumn{2}{c|}{\textbf{Worse-ear (ind.)}} & \multicolumn{2}{c|}{\textbf{Average-ear (ind.)}} & \multicolumn{2}{c|}{\textbf{Both-ear (exch.)}} & \multicolumn{2}{c}{\textbf{Proposed}} \\ \cline{2-9} & Est. (SE) & 95\% CI & Est. (SE) & 95\% CI & Est. (SE) & 95\% CI & Est. (SE) & 95\% CI \\ \hline 

 DASH	&   -0.177 (0.146)	&(-0.463, 0.110)&  -0.229 (0.168) &	(-0.558, 0.099) &  -0.327 (0.170) &	(-0.661, 0.007) & -0.307 (0.169)
& (-0.638, 0.023)
	
\\ 	I(q=2)$\times \text{DASH}$ & 1.210 (0.044) & (1.124, 1.295)&  -0.326 (0.196) & (-0.710, 0.057) &  -0.298 (0.195) & (-0.681, 0.085) & -0.308 (0.195) & (-0.690, 0.074)

\\ I(q=3)$\times \text{DASH}$ &  0.210 (0.254) & (-0.289, 0.708) &  -0.334 (0.297) & (-0.917, 0.248) & -0.384 (0.296) &	(-0.964, 0.197) & -0.343 (0.296) & (-0.923, 0.237)

\\hearing threshold &  0.004 (0.004) & (-0.004, 0.013) & -0.011 (0.006) & (-0.023, 0.000) & -0.091 (0.007) & (-0.105, -0.078) &  -0.062 (0.007) & (-0.075, -0.048)

\\
\bottomrule
\end{tabular}
\end{sideways}
\end{table}

\noindent where $i = 1, \ldots, N$, $j = 1, 2$, $q = 1, 2, \text{ and } 3$ and $Y_{ijq}$ represents the change of hearing threshold from baseline measurement to 3-year follow-up measurement for subject $i$ from ear $j$, and at frequency level $q$. Each subject $i$ has a total of 6 measurements from 2 ears, at 3 frequency levels. The predictor vector $\mathbf{X}_{ijq}$ includes participant level covariate (DASH) and ear-level covariate (baseline hearing threshold at each frequency), as well as potential participant level confounders: age at follow-up (continuous), race (Black v.s. non-Black), body mass index (categorical; $<$25, 25-29, 30-34, 35-30, 40+), physical activity (quintiles), smoking status (categorical, yes/no), history of tinnitus (yes/no), and noise exposure ($\ge$3 hours per week during any decade; yes/no). The DASH scores are scaled so that one unit increase corresponds to increase from 10\% quantile to 90\% quantile. The CHEARS data has seven frequency levels. We categorize them into low (0.5 kHz, 1 kHz, 2 kHz), mid (3 kHz, 4 kHz) and high frequencies (6 kHz, 8 kHz). Outcome at each frequency category is calculated by averaging the outcomes measurements within.

\indent We used mixed effect model to calculate the intra-cluster correlation (ICC), and found a significant moderate ICC among measurements from the same ear (ICC = 0.57, 95\% CI = [0.52, 0.64]) or at the same frequency level (ICC = 0.64, 95\% CI = [0.59, 0.70]). The structure of the correlation matrix also agrees with our assumption shown in Table 1. Table 5 shows the results from four methods: worse-ear (ind.), average-ear (ind.), both-ear (exch.) and the proposed method. In general, the proposed method has smaller SE and narrower 95\% confidence intervals for the coefficient estimates, compared to the other 3 methods. Sample codes is available from the corresponding author.

\section{Discussion}

\indent In this paper, we propose a method that models the within-cluster (participant) correlations when using GEE to estimate the effects of exposures on hearing loss. Our method accounts for ear-level and frequency-level correlations in hearing loss data. Our method leads to a smaller empirical SE comparing to existing methods in the simulation studies, which shows that modelling the correlation matrix in GEE could improve efficiency, especially when estimating the coefficients for covariates corresponding to within-cluster variation (e.g., ear-level covariates). Our method can grant more profound efficiency gain when the within-cluster correlation is moderate to strong, or when comparing to GEE with unstructured correlation matrix. Other data such as eye-sight measurements may have similar correlation structure as hearing loss data, so our method can be applied to other type of studies. Although we focus on a continuous outcome in the simulation study and data example in this paper, the method derivation in Sections 2 and 3 apply for non-continuous outcome through using a non-identity link function $g(\cdot)$. 

\indent We employ GEE1.5 method that estimates correlation coefficients by jointly solving a set of estimating equations, rather than the commonly used GEE1 method that relies on methods of moments \cite{prentice1988}. We implemented GEE1.5 approach as a non-iterative procedure, which is less computationally complex than GEE2 \cite{kalema}. 

\indent Our results also agree with the findings of Sheng et al. (2022) that when the ear-level covariate is present, worse-ear method is typically more biased, especially in estimating ear-level effects; the average-ear estimators may have comparable finite-sample bias to the both-ear estimators, but they are less efficient than the both-ear estimator.

\indent 

\section*{Acknowledgments}

This work was partially supported by the National Institute Health grants R01 DC017717, U01 CA176726 (NHS II) and U01 DC 010811.

\subsection*{Conflict of interest}

The authors declare no potential conflict of interests.

\bibliography{ref.bib}

\newpage 

\appendix

\beginsupplement

\begin{table}[htbp]
\caption{Simulation Results from Scenario 2 (strong correlation)}
\centering
\begin{tabular}{c|l|ccc|ccc}
\toprule
\multirow{2}{2.5em}{} & \multirow{2}{5.5em}{\textbf{\makecell{Method}}}  & \multicolumn{3}{c|}{\textbf{Number of participants = 200}} & \multicolumn{3}{c}{\textbf{Number of participants = 500}} \\ 
\cline{3-8} &
& \makecell{Relative \\ Bias (\%)} & ESE &  CR (\%) & \makecell{Relative \\ Bias (\%)} & ESE & CR (\%)\\
\hline 
 & Worse-ear (ind.) & -103.949 & 3.087 & 92.0 & -98.465 & 1.989 & 89.7\\ 
 & Average-ear (ind.) & 8.856 & 3.076 & 93.4 & -0.648 & 1.922 & 94.3\\
 $\beta_2$ & Both-ear (ind.) & 8.403 & 3.080 & 93.4 & -0.638 & 1.923 & 94.5\\
& Both-ear (exch.) & 7.973 & 3.093 & 93.1 & -0.517 & 1.925 & 94.3\\
& Both-ear (uns.) & -68.921 & 21.392 & 95.8 & -11.528 & 4.096 & 97.2\\
  & Proposed &  8.418 & 3.084 & 93.3 & -0.510 & 1.923 & 94.4\\
\hline
 & Worse-ear (ind.) &  -116.674 & 4.385 & 93.1 & -107.037 & 2.874 & 91.8 \\  
   & Average-ear (ind.) &  9.979 & 2.491 & 94.1 & 1.225 & 1.458 & 96.4\\ 
$\beta_3$ & Both-ear (ind.) &  9.642 & 2.486 & 94.0 & 1.256 & 1.458 & 96.5 \\
 & Both-ear (exch.) & 9.149 & 2.480 & 93.9 & 1.210 & 1.459 & 96.5\\
  & Both-ear (uns.) &  164.408 & 46.126 & 95.5 & 0.437 & 5.479 & 96.9 \\
& Proposed & 9.707 & 2.482 & 93.8 & 1.171 & 1.458 & 96.4\\ 

\hline

 & Worse-ear (ind.) & -112.136 & 0.305 & 1.7 & -113.757 & 0.198 & 0.0 \\  
 & Average-ear (ind.) & 0.443 & 0.295 & 91.4 & -0.756 & 0.188 & 94.2\\ 
$\theta$   & Both-ear (ind.) & 0.133 & 0.286 & 91.6 & -0.492 & 0.187 & 93.9 \\
 & Both-ear (exch.) &  -0.262 & 0.273 & 92.5 & 0.014 & 0.184 & 94.8 \\ 
& Both-ear (uns.) & -183.598 &  39.065 & 91.7 & -21.049 & 6.522 & 95.4 \\
& Proposed & 0.017 & 0.254 & 92.0 & -0.246 & 0.167 & 94.8\\ 

\bottomrule
\end{tabular}
\begin{tablenotes}
\begin{itemize}
    $\;\;\; ^a$ ``SES'' is the empirical SE.\\
    $\;\;\; ^b$ ``ind.'', ``exch.'', and ``uns'' stand for independence, exchangeable and unstructured correlation matrix. \\
    $\;\; ^c$ The coefficient $\beta_2$, $\beta_3$ and $\theta$ correspond to the participant-level exposure, the interaction between participant-level exposure and frequency, and the ear-level covariate, respectively.  
\end{itemize}
\end{tablenotes}
\end{table}

\begin{table}[htbp]
\caption{Simulation Results from Scenario 3 (moderate correlation)}
\centering
\begin{tabular}{c|l|ccc|ccc}
\toprule
\multirow{2}{2.5em}{} & \multirow{2}{5.5em}{\textbf{\makecell{Method}}}  & \multicolumn{3}{c|}{\textbf{Number of participants = 200}} & \multicolumn{3}{c}{\textbf{Number of participants = 500}} \\ 
\cline{3-8} &
& \makecell{Relative \\ Bias (\%)} & \makecell{ESE}&  CR (\%) & \makecell{Relative \\ Bias (\%)} & ESE & CR (\%)\\
\hline 
 & Worse-ear (ind.) & -102.064 & 2.803 & 92.0 & -104.214 & 1.828 & 88.4 \\ 
 & Average-ear (ind.) & 3.999 & 2.936 & 93.0 & 4.700 & 1.846 & 94.8\\
 $\beta_2$ & Both-ear (ind.) & 3.953 & 2.937 & 93.2 & 4.692 & 1.846 & 94.8\\
& Both-ear (exch.) &  3.911 & 2.937 & 93.4 & 4.681 & 1.846 & 94.9\\
& Both-ear (uns.) & -2.073 & 4.650 & 95.4 & 4.754 & 1.910 & 95.7\\
  & Proposed & 3.993 & 2.934 & 93.1  & 4.681 & 1.845 & 94.8\\ 
\hline
 & Worse-ear (ind.) & -100.809 & 3.860 & 94.4 & -110.865 & 2.625 & 93.2 \\  
   & Average-ear (ind.) & -1.788 & 2.849 & 95.4 & 3.832 & 1.795 & 94.8 \\ 
$\beta_3$ & Both-ear (ind.) & -1.507 & 2.850 & 95.4 & 3.889 & 1.797 & 95.0\\
 & Both-ear (exch.) & -1.020 & 2.850 & 95.1 & 4.030 & 1.800 & 95.0\\
  & Both-ear (uns.) &  -15.677 & 6.013 & 94.7 & 4.589 & 3.313 & 95.4\\
& Proposed & -1.298 & 2.849 & 95.2 & 3.970 & 1.797 & 94.9 \\ 

\hline

 & Worse-ear (ind.) & -116.667 & 0.284 & 0.5 & -118.458 & 0.179 & 0.0\\  
 & Average-ear (ind.) & 1.439 & 0.273 & 92.2 & -0.020 & 0.168 & 94.4\\ 
$\theta$   & Both-ear (ind.) & 2.015 & 0.268 & 92.3 & -0.063 & 0.167 & 95.0\\
 & Both-ear (exch.) & 2.277 & 0.260 & 92.3 & -0.265 & 0.165 & 94.9 \\ 
& Both-ear (uns.) &  23.439 & 2.673 & 91.0 & -0.601 & 0.389 & 94.0\\
& Proposed &  1.880 & 0.251 & 92.8 & -0.181 & 0.157 & 95.1\\ 

\bottomrule
\end{tabular}
\begin{tablenotes}
\begin{itemize}
    $\;\;\; ^a$ ``SES'' is the empirical SE. \\
    $\;\;\; ^b$ ``ind.'', ``exch.'', and ``uns'' stand for independence, exchangeable and unstructured correlation matrix.\\
    $\;\; ^c$ The coefficient $\beta_2$, $\beta_3$ and $\theta$ correspond to the participant-level exposure, the interaction between participant-level exposure and frequency, and the ear-level covariate, respectively.  
\end{itemize}
\end{tablenotes}
\end{table}

\begin{table}[htbp]
\caption{Simulation Results from Scenario 4 (weak correlation)}
\centering
\begin{tabular}{c|l|ccc|ccc}
\toprule
\multirow{2}{2.5em}{} & \multirow{2}{5.5em}{\textbf{\makecell{Method}}}  & \multicolumn{3}{c|}{\textbf{Number of participants = 200}} & \multicolumn{3}{c}{\textbf{Number of participants = 500}} \\ 
\cline{3-8} &
& \makecell{Relative \\ Bias (\%)} & \makecell{ESE}&  CR (\%) & \makecell{Relative \\ Bias (\%)} & ESE & CR (\%)\\
\hline 
 & Worse-ear (ind.) & -97.526 & 2.505 & 92.1 & -95.869 & 1.702 & 88.3 \\ 
 & Average-ear (ind.) & -3.721 & 2.879 & 93.5 & 6.008 & 1.750 & 95.6 \\
 $\beta_2$ & Both-ear (ind.) & -3.835 & 2.879 & 93.5 & 5.946 & 1.752 & 95.8\\
& Both-ear (exch.) & -4.003 & 2.879 & 93.6 & 5.901 & 1.753 & 95.8\\
& Both-ear (uns.) &  -4.987 & 2.925 & 94.3 & 5.919 & 1.793 & 95.1\\
  & Proposed & -3.873 & 2.878 & 93.5 & 5.921 & 1.752 & 95.8 \\ 
\hline
 & Worse-ear (ind.) & -85.339 & 3.616 & 93.5 & -94.407 & 2.286 & 93.0\\  
   & Average-ear (ind.) &  -4.460 & 3.411 & 95.4 & 1.298 & 2.182 & 94.9\\ 
$\beta_3$ & Both-ear (ind.) & -4.439 & 3.408 & 95.4 & 1.419 & 2.182 & 95.1\\
 & Both-ear (exch.) & -4.470 & 3.404 & 95.3 & 1.463 & 2.182 & 95.1\\
  & Both-ear (uns.) &  -5.538 & 4.585 & 96.4 & 3.145 & 2.895 & 95.4\\
& Proposed & -4.429 & 3.407 & 95.4 & 1.405 & 2.182 & 95.1\\ 

\hline

 & Worse-ear (ind.) & -122.787 & 0.261 & 0.0 & -123.388 & 0.168 & 0.0 \\  
 & Average-ear (ind.) & 0.259 & 0.247 & 92.1 & 0.268 & 0.162 & 94.1 \\ 
$\theta$   & Both-ear (ind.) & 0.397 & 0.247 & 92.5 & 0.081 & 0.162 & 94.5\\
 & Both-ear (exch.) &  0.506 & 0.246 & 93.4 & 0.008 & 0.162 & 95.5\\ 
& Both-ear (uns.) &  -1.900 & 0.444 & 92.6 & 0.280 & 0.151 & 94.5\\
& Proposed & 0.398 & 0.242 & 92.5 & 0.077 & 0.160 & 94.8 \\ 

\bottomrule
\end{tabular}
\begin{tablenotes}
\begin{itemize}
    $\;\;\; ^a$ ``SES'' is the empirical SE. \\
    $\;\;\; ^b$ ``ind.'', ``exch.'', and ``uns'' stand for independence, exchangeable and unstructured correlation matrix. \\
    $\;\; ^c$ The coefficient $\beta_2$, $\beta_3$ and $\theta$ correspond to the participant-level exposure, the interaction between participant-level exposure and frequency, and the ear-level covariate, respectively.  
\end{itemize}
\end{tablenotes}
\end{table}

\begin{table}[htbp]
\caption{Relative efficiency for $\theta$ under varying strength of correlation (Number of participants = 500)}
\centering
\begin{tabular}{l|cccc}
\toprule
  \multirow{2}{5.5em}{\textbf{\makecell{Method}}}  & \multicolumn{4}{c}{\textbf{Relative Efficiency}}  \\ 
\cline{2-5} &
 \makecell{Scenario 1 \\ (very strong)} & \makecell{Scenario 2 \\ (strong)}& \makecell{Scenario 3 \\ (moderate)} & \makecell{Scenario 4 \\ (weak)}\\
\hline

  Average-ear (ind.)& 0.807  & 0.888 &  0.935&  0.988\\ 
 
Both-ear (ind.) &  0.827 & 0.893  & 0.940 &  0.988\\ 

  Both-ear (exch.) & 0.867  & 0.908 & 0.952 & 0.988 \\ 

 Both-ear (uns.) &  0.277 &  0.026 &  0.404& 1.06  \\

\bottomrule
\end{tabular}
\begin{tablenotes}
\begin{itemize}
     $\;\;\; ^a$ Relative efficiency is the ratio of empirical SE of our proposed method to that of other methods. \\
     $\;\;\; ^b$ ``ind.'', ``exch.'', and ``uns'' stand for independence, exchangeable and unstructured correlation matrix. \\
     $\;\;\; ^c$ The coefficient $\theta$ corresponds to the ear-level covariate.  
\end{itemize}
\end{tablenotes}
\end{table}

\end{document}